\documentclass[conference]{IEEEtran}
\IEEEoverridecommandlockouts
\usepackage{cite}
\usepackage{amsmath,graphicx,url,times,booktabs,tabularx,amsfonts,siunitx}
\usepackage{algorithmic}
\usepackage{graphicx}
\usepackage{textcomp}
\usepackage{xcolor}
\usepackage{url}
\usepackage{multirow}
\usepackage{siunitx}
\usepackage{soul}

\usepackage{microtype}
\usepackage{color}
\usepackage[colorlinks,linkcolor=blue,]{hyperref}
\usepackage[caption=false]{subfig}
\usepackage{subfig}

\def\BibTeX{{\rm B\kern-.05em{\sc i\kern-.025em b}\kern-.08em
    T\kern-.1667em\lower.7ex\hbox{E}\kern-.125emX}}
\begin{document}

\let\OLDthebibliography\thebibliography
\renewcommand\thebibliography[1]{
  \OLDthebibliography{#1}
  \setlength{\parskip}{0pt}
  \setlength{\itemsep}{4.4pt plus 0.1ex}
}

\title{Deep Neural Decision Forest for Acoustic \\Scene Classification}

\author{\IEEEauthorblockN{1\textsuperscript{st} Given Name Surname}
\IEEEauthorblockA{\textit{dept. name of organization (of Aff.)} \\
\textit{name of organization (of Aff.)}\\
City, Country \\}}

\author{
      \IEEEauthorblockN{
      Jianyuan Sun$^{1,3,*}$,
      Xubo Liu$^{1,*}$,
      Xinhao Mei$^{1}$,
      Jinzheng Zhao$^{1}$, 
      Mark D. Plumbley$^{1}$,\\
      Volkan Kılıç$^{2}$,
      Wenwu Wang$^{1}$
     }\thanks{* The first two authors contributed equally to this work.}
     \\
      \IEEEauthorblockN{$^1$Centre for Vision, Speech and Signal Processing (CVSSP), University of Surrey, UK}
      \IEEEauthorblockN{$^2$Department of Electrical and Electronics Engineering, Izmir Katip Celebi University, Turkey}
      \IEEEauthorblockN{$^3$College of Computer Science and Technology, Qingdao University, China}      
      }
\maketitle

\begin{abstract}
Acoustic scene classification (ASC) aims to classify an audio clip based on the characteristic of the recording environment. In this regard, deep learning based approaches have emerged as a useful tool for ASC problems. Conventional approaches to improving the classification accuracy include integrating auxiliary methods such as attention mechanism, pre-trained models and ensemble multiple sub-networks. However, due to the complexity of audio clips captured from different environments, it is difficult to distinguish their categories without using any auxiliary methods for existing deep learning models using only a single classifier. In this paper, we propose a novel approach for ASC using deep neural decision forest (DNDF). DNDF combines a fixed number of convolutional layers and a decision forest as the final classifier. The decision forest consists of a fixed number of decision tree classifiers, which have been shown to offer better classification performance than a single classifier in some datasets. In particular, the decision forest differs substantially from traditional random forests as it is stochastic, differentiable, and capable of using the back-propagation to update and learn feature representations in neural network. Experimental results on the DCASE2019 and ESC-50 datasets demonstrate that our proposed DNDF method improves the ASC performance in terms of classification accuracy and shows competitive performance as compared with state-of-the-art baselines.

\end{abstract}

\section{Introduction}
Acoustic scene classification (ASC) has attracted much attention in the fields of Audio and Acoustic Signal Processing (AASP) \cite{2020Pyramidal,2020A}, as shown in the Detection and Classification of Acoustic Scenes and Events (DCASE) challenges held in recent years, where several benchmark datasets have been introduced. The ASC task focuses on recognizing the audio clips in terms of the type of acoustic environment where they were captured. They are useful in applications such as health care \cite{2017Snore, liu2020speech}, and security surveillance \cite{2020A}.

In past few years, many methods have been developed to address the challenges of ASC, and to improve the performance of ASC. The classical ASC methods tend to employ hand-crafted features, including the Mel Frequency Cepstral Coefficients (MFCCs) \cite{2016CP, liu2021conditional}, spectrogram and log-mel filter banks \cite{0DCASE}, and to train the well-known classifier, such as support vector machine (SVM) \cite{2014Large} and decision trees \cite{P16}. However, theoretical and algorithmic advances together with the increasing capability in computer processing have led to the emergence of more sophisticated methods in artificial intelligence. A representative method for ASC tasks is deep learning which offers superior performance in handling a large number of features. The audio sequences are first converted to a 2-dimensional representation with the time-frequency methods, including log-mel spectrogram, wavelet transform, and short-time Fourier transform, as an input of a deep learning system. In particular, CNNs \cite{2016CP, P17} and recurrent neural networks based methods were shown to provide state-of-the-art performance on some ASC datasets \cite{0ACOUSTIC}. Moreover, the CNN variations such as VGG and ResNet are applied to learn ASC representation \cite{Ma18,ZhangHS20}. 

To further improve the classification performance, deep learning based approaches are extended with some auxiliary methods, i.e., attention mechanism, pre-trained models and ensemble sub-networks. Ding et al. proposed an ensemble system of the CNN and Gaussian Mixture Model (GMM) based on learned features to improve the classification performance \cite{Ense19}. Sugagara et al. improved the accuracy by the ensemble of ResNet-based models and data augmentation methods, such as mixup, time-shifting, and SpecAugment \cite{EnsemRE21}. Han et al. introduced the attention mechanism to improve deep CNN performance \cite{liang2019acoustic}. Moreover, Huang et al. improved the deep CNN using spatial-temporal attention pooling \cite{zhenyi2019acoustic}. Bilot et al. proposed a fusion system that uses multi-layer perceptron (MLP) to get the final results from the initial class label probability predictions \cite{bilot2019acoustic}. In addition, Wang et al. extracted spectral features, i.e., the MFCC of audio files and used a custom-designed CNN for the recognition \cite{wang2019acoustic}. 

Conventional methods focus on learning better feature representations by proposing diverse network structures, which usually use the softmax classifier as the final layer. However, due to the complexity of audio clips collected from different environments, it is difficult for existing deep learning models using only a single classifier, i.e., Softmax, to distinguish audio clip categories. In machine learning, there are various well-known classifiers, such as support vector machine (SVM) \cite{2014Large} and random forests \cite{2001Random}, which show an outstanding classification or prediction performance. In particular, the traditional random forests algorithm has achieved great success in practical applications, which is a typical ensemble method combining a fixed number of decision trees \cite{2001Random}. However, combining deep learning with the random forests method has received little attention due to the limitations associated with the local optimal strategy, i.e., calculating node features and split thresholds using the Gini index or information gain rate, which results in the challenge in performing back-propagation for combined models \cite{li1984classification, KontschiederFCB15}. Kontschieder et al. addressed this limitation by introducing DNDFs approach that unifies random forests using the representation learning with deep CNNs, which enables end-to-end training \cite{KontschiederFCB15}. It was reported that DNDF shows state-of-the-art performance as compared to its machine learning counterparts \cite{2014Neural, ZhangBOK21}.   

In this paper, we investigate the performance of DNDF in ASC. The existing deep learning methods focus on learning good feature representations using the attention mechanism and ensemble sub-networks. Most methods routinely use the softmax classifier as the final layer which may not be sufficient for classifying audio clips of diverse categories from complex environments. Therefore, in our work, we employ a decision forest classifier instead of the softmax as the final predictor. 
Our work is the first attempt to apply the DNDF for the ASC task. Experiments show that the DNDF method achieves competitive results with the existing state-of-the-art model that uses a pre-trained model. Moreover, the DNDF method obtains better performance as compared to most deep CNN models with the attention mechanism and ensemble sub-networks.

The remainder of this paper is organized as follows. The next section introduces our proposed method. Section \ref{sec:Experiments} presents experimental setup. Section \ref{sec:Results} shows experimental results on the DCASE2019 development ASC subtask A and ESC-50 dataset. Conclusions are given in Section \ref{sec:conclusion}.

\section{Methods}
\subsection{Deep Neural Decision Forests (DNDFs)}
\begin{figure}[ht]
 \centering
\begin{minipage}[b]{1\linewidth}   
 \centerline{\includegraphics[height=4.7cm,width=6.4cm]{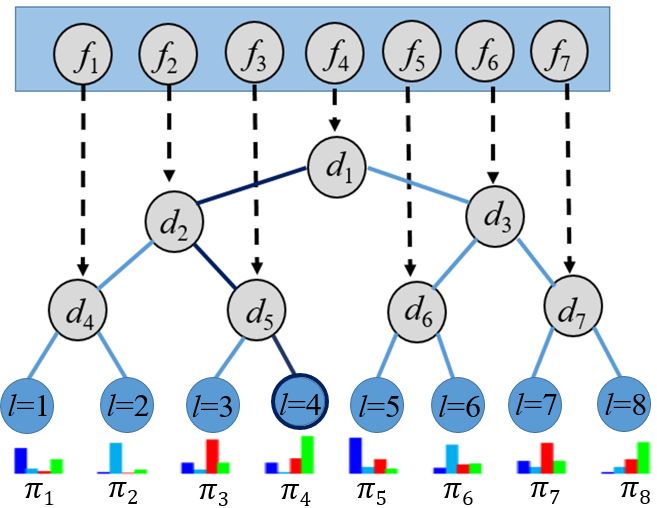}}
\end{minipage}
\caption{The architecture of the DNDF model. In this DNDF, seven embedding functions $f_n$ ($n = 1,\ldots, 7$) are provided by the final layer of CNN, which is a real-value function depending on the sample and the parameter $\theta$. Each output of $f_n$ determines the decision function $d_n$ of each node for the decision trees. Each prediction node at the bottom of the decision tree has the probability distribution $\pi_l$ for each class.}
\label{fig:DNDF}
\end{figure}

DNDF is a type of CNN that replaces the softmax layer with decision forests, consisting of several decision trees. Given a classification dataset with input and (finite) output space $X$ and $Y$. A decision tree is a binary tree consisting of decision nodes and prediction nodes. Here, the symbol $\mathcal{N}$ is used to denote the decision node index of a decision tree. $\mathcal{L}$ denotes the set of the prediction node indices $\{1,...,L\}$. Each prediction node $l\in \mathcal{L}$ has a probability distribution $\pi_{l}$ over $Y$. Each decision node $n\in \mathcal{N}$ is assigned a decision function $d_{n}(\cdot;\theta)$, where the parameter $\theta$ from the CNN is used to update the feature representation. Moreover, the embedding function $f_n(\cdot;\theta) : X \rightarrow R$ is defined in CNN, which will determine the action of the decision function $d_n(\cdot;\theta)$ of the decision trees. Fig. \ref{fig:DNDF} shows the architectural structure of DNDF, showing how decision nodes can be implemented by using the output of the final layer of CNN. To easy understanding, we only show the example of building a single decision tree in DNDF by using a fixed number of CNN embedding functions. 

In DNDF, the fully-connected and convolutional layers are the same as those in a general CNN. The feature representations learned by the fully-connected layer are used as the tree node of the decision trees in decision forests. Therefore, CNN nodes share the same parameter $\theta$ that is used to update the feature representation of CNN as the tree nodes. The decision function of each decision node $d_n(\cdot;\theta)$ is defined as follows  
\begin{equation}\label{eq0}
d_n(x;\theta) = \sigma(f_n(x;\theta)),
\end{equation}
where $x\in X$ is the sample input, and $f_n(\cdot;\theta)$ is a real-valued function depending on the sample and the parameters $\theta$, which can be regarded as the linear output unit of the neural network node. When a sample $x\in X$ arrives at a node, whether it goes to the left or right subtree of this node is determined by the output of $d_n(x;\theta)$. In DNDF, a sample arrives from a tree node to a leaf node via stochastic routing. The routing function \mbox{$\mu_l(x|\theta)$} is defined as follows,
\begin{equation}
  \mu_l(x|\theta)=\prod_{n\in N}d_n(x;\theta)^{\swarrow}\bar{d_{n}}(x;\theta)^{\searrow},
\end{equation}
where $\bar{d}_{n}(x;\theta)=1-d_{n}(x;\theta)$. $d_{n}^\swarrow$ represents the route from the current node to the left and $l$ is the leaf node. If a sample goes to reach the leaf node $l=4$:
\begin{equation}
  \mu_{(l=4)}=d_1(x)\bar{d_{2}}(x)\bar{d_{5}}(x).
\end{equation}

Under the stochastic routing strategy, a sample arrives at a leaf node $l$, the related tree prediction is determined by the class label distribution $\pi_l$. $\pi_{l^{y}}$ represents the probability of sample reaching leaf node $l$ to take on class $y$. The final prediction for a sample that takes on a class $y$ is the average probability results of reaching the leaf node. The definition is 
\begin{equation}\label{eq2}
  P[y|x,\theta,\pi]=\sum_{l\in \mathcal{L}}\pi_{l^{y}}\mu_{l}(x|\theta).
\end{equation}

For the decision forests, it is an ensemble of several decision trees $\mathbb{F}={T_{1},...,T_{k}}$. The final prediction of decision forests for a sample $x$ is the average output of each tree, that is,   
\begin{equation}\label{eq3}
 P_{F}[y|x]=\frac{1}{k}\sum_{h=1}^{k}P_{T_h}[y|x].
\end{equation}

\subsection{Application of DNDF in ASC}
\begin{figure}[ht]
 \centering
\begin{minipage}[b]{1\linewidth}   
 \centerline{\includegraphics[height=6.4cm,width=8.3cm]{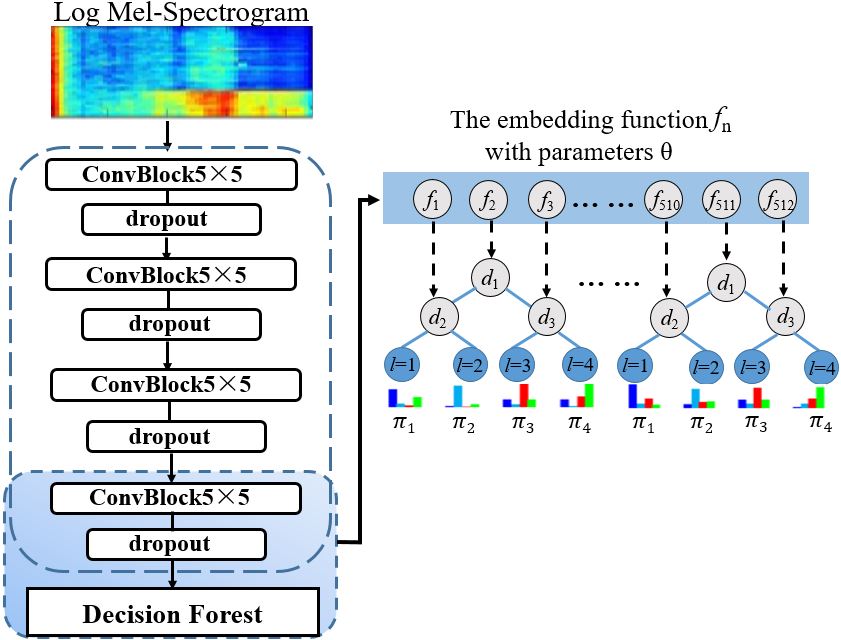}}
\end{minipage}
\caption{The architecture of the DNDF model for ASC. $512$ embedding functions $f_n$ ($n = 1,\ldots, 512$) are provided by the final layer of CNN4. Each output of $f_n$ determines the decision function $d_n$ of each node for the decision trees. Each prediction node at the bottom of the decision tree has the probability distribution $\pi_l$ for each acoustic scene class.} 
\label{fig:DNDF2}
\end{figure}

We proposed a DNDF for the task of ASC. Specifically, DNDF consists of two parts, i.e., CNN and decision forests. In particular, we design a CNN4 with four convolutional blocks. Each convolutional block has one convolutional layer with a kernel size of $5\times5$. After each convolutional layer, batch normalization and ReLU are used. The channel numbers of each convolutional block are $64, 128, 256, 512$, respectively. Moreover, an average pooling layer with kernel size $2\times2$ is employed between two neighbouring blocks for down-sampling. The decision forest is used after the fourth convolutional block. Here, DNDF takes log Mel-spectrogram features of the acoustic clips as the inputs. Fig. \ref{fig:DNDF2} shows the proposed architecture of DNDF for ASC. In this DNDF, $512$ embedding functions $f_n(\cdot;\theta)$ ($n = 1,\ldots, 512$) are provided by the final layer of CNN4. Where, the parameter $\theta$ is used to update the feature representation. Each output of $f_n(\cdot;\theta)$ determines the decision function $d_n(\cdot;\theta)$ of each node for the decision trees, i.e, Eq. (\ref{eq0}). Each prediction node at the bottom of the decision tree has the probability distribution $\pi_l$ for each acoustic scene class, i.e., Eq. (\ref{eq2}) and Eq. (\ref{eq3}).

\subsection{Learning Procedure of DNDF}
In DNDF, given a training dataset $\mathcal{T}\subset X \times Y$, we start with random initialization of the common parameter $\theta$ of the decision trees and convolutional network. Furthermore, an iterative learning procedure is performed with a predefined number of epochs. Given the value of $\theta$, the estimation of the prediction node parameters $\pi$ are obtained, setting the initial value in each leaf node as $\pi_{l^{y}}^{(0)}=|Y|^{-1}$. To estimate and update parameters $\theta$ and $\pi$, we search for the minimizers of the following empirical risk function, i.e., 
\begin{equation}\label{eq1}
R(\theta,\pi;\mathcal{T})=\frac{1}{|\mathcal{T}|}\sum_{(x,y)\in{\mathcal{T}}}L(\theta,\pi;x,y),
\end{equation}
where $L(\theta,\pi;x,y)=-log(P_{T}[y|x, \theta, \pi])$ denotes the log-loss term for the training sample $(x,y)\in{\mathcal{T}}$. 
Moreover, we divide the training data into mini-batches. Then, we perform stochastic gradient descent (SGD) to update the parameter $\theta$ by minimizing the empirical risk function based on each mini-batch. This learning process is given in Algorithm 1.
\begin{table}[ht]
\centering
 \begin{tabular}{lccl}
  \toprule
  Algorithm 1: The parameter learning process of DNDF  \\
  \midrule
  \qquad \textbf{Require}: Given training set $\mathcal{T}$, epochs $=K$ \\
  \qquad Initialization parameter $\theta$ \\
  \qquad \textbf{For each} Epoch $i\in\{1,...,K\}$ \textbf{do} \\
  \qquad   \ \ Using an iterative scheme to compute $\pi$  \\
  \qquad   \ \ Split $\mathcal{T}$ into a fixed number of random mini-batches \\
  \qquad   \ \ \ \ \textbf{For each} mini-batch \textbf{do}  \\
  \qquad     \ \ \ \ \ \ Using SGD update $\theta$ by minimizing Eq. (\ref{eq1}) \\
  \qquad   \ \ \ \ \textbf{end for} \\
  \qquad \textbf{end for} \\
  \bottomrule
 \end{tabular}
\end{table}

\section{Experiments} 
\subsection{Dataset} 
To evaluate the performance of DNDF in ASC, the DCASE 2019 development ASC subtask A dataset and ESC-50 environmental sound classification dataset are used in our experiments. DCASE 2019 is an extension of the DCASE 2018 TUT Urban ASC dataset with 10-second-long clips. There are 10 acoustic scenes in DCASE 2019, including bus, metro, metro\_station, park, public\_square, street\_pedestrian, shopping\_mall, tram, street\_traffic and airport. The DCASE 2019 ASC dataset contains 9185 audio clips for training and 4185 clips for testing, sampled at \SI{48000}{\Hz}. The ESC-50 dataset contains 2000 environmental audio clips each of 5-seconds, sampled \SI{41000}{\Hz}, from 50 semantic classes. 

\subsection{Audio Processing and Augmentation}
The original audio clip is converted to 64-dimensional log Mel-spectrogram by using the short-time Fourier transform with a frame size of \num{1024} samples, a hop size of \num{320} samples, and a Hanning window. In addition, SpecAugment is used for data augmentation \cite{park2019specaugment}.

\subsection{Baseline}
Our proposed method does not use auxiliary techniques such as attention mechanism and pre-trained model. Therefore, we choose the CNN-based variations as well as some CNN methods that use an attention mechanism, pre-trained and ensemble multiple sub-network as the baseline methods.

For the DCASE 2019 ASC dataset, we choose CDNN\_CRNN \cite{pham2019cdnn}, Attention\_CNN \cite{liang2019acoustic}, HPSS\_MFCC\_CNN \cite{zhenyi2019acoustic}, MIL\_CNN \cite{bilot2019acoustic} and MFCC\_CNN \cite{wang2019acoustic} for the comparison. The CDNN\_CRNN \cite{pham2019cdnn} model is a joint learning model based on a Convolutional Deep Neural Network and Convolutional Recurrent Neural Network. The Attention\_CNN \cite{liang2019acoustic} model improves the performance of deep CNN by introducing an attention mechanism. The HPSS\_MFCC\_CNN \cite{zhenyi2019acoustic} uses deep CNN with spatial-temporal attention pooling. Moreover, the mixup data augmentation technique is employed to improve the classification performance further. Inspired by the multiple instant learning, the MIL\_CNN \cite{bilot2019acoustic} model uses MLP to get the final results from the initial class label probability predictions. The MFCC\_CNN \cite{wang2019acoustic} model extracts the MFCC feature from audio files and uses the CNN with four-layer convolution and two-layer fully connected layer for classification.   

For the ESC-50 dataset, we compared existing algorithms including WELACNN \cite{KumarKF18}, ACLNet \cite{abs-1811-06669}, ENSCNN \cite{nanni2021ensemble} and ACDNet \cite{mohaimenuzzaman2021environmental}. The WELACNN model \cite{KumarKF18} uses the transfer learning technique to learn knowledge from weakly labeled audio data based on a CNN. The ACLNet \cite{abs-1811-06669} introduces an efficient CNN architecture by using data augmentation and regularization. The ENSCNN \cite{nanni2021ensemble} ensembles classifiers by combining six data augmentation techniques and four signal representations to train five pre-trained CNNs. The ACDNET \cite{mohaimenuzzaman2021environmental} uses a large deep CNN as a pipeline of a network for edge devices with resource constraints. The work \cite{gong21binterspeech} is a state-of-the-art attention-based CNN model with transformer encoder, pre-trained on AudioSet \cite{GemmekeEFJLMPR17}.

\subsection{Training Procedure} \label{sec:Experiments}
The DNDF model is trained by employing the Adam optimizer with a learning rate of $0.001$. Moreover, the batch size is set to $150$, the number of epochs is $500$, the decision trees depth is $10$, and the number of decision trees is $100$. In particular, for the ESC-50 dataset, to ensure the same settings as the comparison methods, we train a DNDF model with 5-fold cross-validation and report the average classification accuracy. For the comparison methods, we do not perform the training and testing processes. The accuracy results of the comparison methods are all from the original public results.   

\subsection{Results} \label{sec:Results}
The same evaluation metric adopted in subtask A of the DCASE 2019 is used, i.e., the classification accuracy. Table \ref{table:tabresults1} and Table \ref{table:DNDF3} show the classification results of DNDF and baseline methods on DCASE 2019 ASC subtask A and ESC-50 datasets, respectively.

\begin{table}[ht]
\centering
\caption{Mean classification accuracy of the compared methods on the DCASE2019 development ASC subtask A. The best result is shown in boldface.}
\setlength{\tabcolsep}{6mm}
\begin{tabular}[\linewidth]{c c c} 
 \hline
 Model & DCASE2019 \\ 
 \hline
 CDNN\_CRNN\cite{pham2019cdnn} & 73.70 \\ 
 Attention\_CNN\cite{liang2019acoustic} & 70.70 \\
 HPSS\_MFCC\_CNN\cite{zhenyi2019acoustic} & 73.90 \\
 MIL\_CNN\cite{bilot2019acoustic} & 72.30 \\
 MFCC\_CNN\cite{wang2019acoustic} & 73.50 \\
 DCASE2019\_baseline &  63.30   \\
 DNDF (ours) & \textbf{75.90}   \\
 \hline
\end{tabular}
\label{table:tabresults1} 
\end{table}

\begin{table}[ht]
\centering
\caption{Mean classification accuracy of the compared methods on the ESC-50 dataset. The best accuracy is shown in boldface.}
\setlength{\tabcolsep}{9mm}
\begin{tabular}[\linewidth]{c c c} 
 \hline
 Model & ESC-50 \\ 
 \hline
 WELACNN\cite{KumarKF18} & 83.50  \\
 ACLNet\cite{abs-1811-06669} & 85.65  \\
 ENSCNN\cite{nanni2021ensemble} & 88.65  \\
 ACDNet\cite{mohaimenuzzaman2021environmental} & 87.10 \\
 SOTA\cite{gong21binterspeech} & \textbf{95.70} \\
 DNDF (ours) & \underline{88.90} \\
 \hline
\end{tabular} 
\label{table:DNDF3} 
\end{table}

\begin{table}[ht]
\centering
\caption{The classification accuracy of DNDF with different numbers of decision trees on the DCASE2019 development ASC subtask A and the ESC-50 dataset. The best accuracy is shown in boldface.}
\setlength{\tabcolsep}{6mm}
\begin{tabular}[\linewidth]{c | c c c} 
 \hline
 Number of trees   & DCASE2019 & ESC-50  \\ 
 \hline
 5               & 72.80 & 84.80    \\
 10              & 73.80 & 87.50    \\
 20              & 74.40 & 88.50    \\
 50              & 74.50 & 88.60    \\
 80              & 75.70 & 87.70    \\
 100             & \textbf{75.90} & \textbf{88.90}   \\
 \hline
 \#Tree depth      & 10   & 10        \\
 \#Batch size      & 150  & 150       \\
 \hline
\end{tabular} 
\label{table:tabresults2} 
\end{table}

Experimental results show that the DNDF method outperforms the baseline CNN-based methods. In particular, DNDF obtains the best accuracy as compared with the CNN-based model with the attention mechanism \cite{liang2019acoustic,zhenyi2019acoustic} and ensemble multiple sub-network method \cite{bilot2019acoustic,nanni2021ensemble} as the decision forest can guide the representation learning in lower layers of deep CNN. Moreover, the CNN can gradually learn a good feature representation based on the decision forest prediction results. At the same time, the results also demonstrate that DNDF with decision forest as the final classifier can have satisfactory classification performance. However, it is worth noting that the performance of DNDF does not outperform the SOTA model, because the SOTA method uses the pre-trained model that pre-trains on the Audioset based on the transformer encode model \cite{gong21binterspeech}.

In addition, to investigate the effect of the number of decision trees on the performance of DNDF, we evaluate the use of different numbers of the decision trees, and observe the change in the classification accuracy of DNDF. The results are shown in Table \ref{table:tabresults2}. It can be found that the classification accuracy increases gradually with the number of decision trees. These results illustrate that DNDF is robust with the number of trees. 

\section{Conclusion} \label{sec:conclusion}
In this paper, we presented a DNDF model for the ASC by combining the convolution neural network and decision forest. The traditional random forests have a rich and successful history in machine learning and computer vision. However, the traditional random forests and neural networks cannot be learned together in an end-to-end way due to the non-differentiability of the traditional random forests for the parameters of neural networks. The decision forest in DNDF differs from the conventional random forests because it is stochastic and differentiable. Therefore, the decision forest can learn and organize the feature representation of the deep neural networks. To the best of our knowledge, our work is the first attempt to use the DNDF to solve the ASC task. Experiments demonstrated that the DNDF method achieves competitive results with the existing state-of-the-art model which, however, uses a pre-trained model built on large scale training data. Moreover, the DNDF method obtains better performance than existing deep CNN models with the attention mechanism and ensemble multiple sub-networks.

\section*{Acknowledgment} 
This work is partly supported by a Newton Institutional Links Award from the British Council and the Scientific and Technological Research Council of Turkey (TUBITAK), titled ``Automated Captioning of Image and Audio for Visually and Hearing Impaired" (Grant numbers 623805725 and 120N995), a grant EP/T019751/1 from the Engineering and Physical Sciences Research Council (EPSRC), and a PhD scholarship from the University of Surrey, and a Research Scholarship from the China Scholarship Council. 

\bibliographystyle{IEEEtran}
\bibliography{refs}

\end{document}